\documentstyle[aps,twocolumn]{revtex}
\begin{document}
\voffset 0.8truecm
\title{
Cloning of symmetric d-level photonic states in physical systems
}
\author{
Heng Fan$^a$, Gregor Weihs$^b$, Keiji Matsumoto$^a$, and  Hiroshi
Imai$^a$}
\address{
$^a$Quantum computing and information project,
ERATO, \\
Japan Science and Technology Corporation,\\
Daini Hongo White Bldg.201, Hongo 5-28-3, Bunkyo-ku, Tokyo 133-0033, Japan.\\
$^b$Ginzton Laboratory, Stanford University, Stanford, CA 94305, USA. \\
} \maketitle

\begin{abstract}
Optimal procedures play an important role in quantum information.
It turns out that some naturally occurring processes like
emission of light from an atom can realize optimal
transformations. Here we study how arbitrary symmetric states of
a number of d-level systems can be cloned using a multilevel
atomic system. It is shown that optimality is always ensured even
though the output number of systems is probabilistic.
\end{abstract}

\pacs{03.67.Lx, 03.65.Ta, 32.80.Qk}

The no-cloning theorem, a well known result in quantum
theory\cite{WZ}, states that no unitary transformation can copy
exactly two distinct, nonorthogonal pure states. Also it has been
extended to other cases\cite{BCFJ,KI} such as a no-broadcast
theorem for noncommuting mixed states and no-cloning theorem for
entangled states. However, the no-cloning theorem does not forbid
imperfect cloning. Only in recent years it has been shown that it
is in principle possible to construct a universal quantum cloning
machine (UQCM). This machine will clone arbitrary pure states
equally well\cite{BH}, where the fidelity is the measure of
cloning quality. The UQCM was later proved to be optimal
\cite{BDEF} and is a very nice example of a device that conserves
quantum information. A generalization from 1 to 2 cloning to the
general $N$ to $M$ cloning transformation was proposed by Gisin
and Massar\cite{GM}. Subsequently the cloning of d-level quantum
systems was studied first by completely positive maps\cite{W,KW},
and later by unitary transformations\cite{BH1,FMW,FMWHW}. The
relation between universal cloning and quantum estimation theory
\cite{DBE} was studied in \cite{BEM}, and the upper bound of
cloning for the input in symmetric subspace was obtained. Very
recently general cloning transformations with input in the
symmetric subspace were proposed and proved to be
optimal\cite{FMWHW}.

A widespread view on quantum cloning assumes that it must be
realized by quantum networks\cite{BBHB}. Thus it came a little bit
surprising that stimulated emission of photons can realize the
UQCM \cite{SWZ,KSW} without further engineering, and the
corresponding fidelity is optimal. This provides an easy way to
realize a UQCM and to achieve the upper bound of the cloning
fidelity. In this scheme, it has been shown that certain types of
three-level atoms can be used to optimally clone quantum
information that is encoded as an arbitrary superposition of
excitations in photonic modes. These photonic modes are
associated with corresponding atomic transitions which mediate
the cloning. The universality of the cloning is ensured in the
scheme by requiring symmetry under general unitary transformations
of the system to be cloned (see Ref.~\cite{SWZ,KSW} for detailed
arguments). It should be noted that quantum cloning of orthogonal
qubits proposed in \cite{GP} and extended by Fiurasek et al can
also be realized by the scheme of
stimulated emission of photons \cite{FIMC}.

The purpose of our work is to show that it is possible to
generalize the scheme introduced in \cite{SWZ,KSW} in two
directions. It will be shown first that we can allow for a wider
set of input states and second that the scheme naturally extends
to the cloning of d-level systems. We shall first briefly
introduce the qubit case extended to arbitrary input states in the
Bose subspace. The qubits will be composed by a particle being in
either of two bosonic modes, e. g. two polarization modes. With
these extended initial states, we provide another way to prove
that this cloning scheme is universal.

Then, we shall continue by studying the cloning of states in
d-dimensional Hilbert space. A generalized Hamiltonian with
d+1-level quantum systems as the cloning medium will be studied,
and the optimal cloning transformation for arbitrary symmetric
states in d-dimensional Hilbert space is obtained as a result. 

Finally we will briefly discuss the experimental issues for
d-dimensional quantum cloning. We remark that our results imply
that whenever a cloning system can be represented by bosonic
operators, we can clone arbitrary states by this UQCM with a
fidelity that achieves its upper bound.

We first briefly review the quantum cloning scheme proposed in
\cite{SWZ,KSW}. The cloning device is an inverted medium that can
spontaneously emit photons of any polarization with the same
probability. This symmetry property will ensure that the cloning
transformation induced by the inverted medium is universal. For
the qubit case, the initial state medium should consist of a
population inverted ensemble of three-level $\Lambda$ atoms. The
system has two degenerate ground states $|g_1\rangle $ and
$|g_2\rangle $ and an excited level $|e\rangle $.\footnote{On
could also start out with $V$-type three-level systems but in
$\Lambda$-systems the initial state turns out to be much simpler.}
The ground states are coupled to the excited state by two modes
of the electromagnetic field $a_1$ and $a_2$, respectively. The
interaction between field and medium is described by the
Hamiltonian
\begin{eqnarray}
H=\gamma \left( a_1\sum _{k=1}^N|e^k\rangle \langle g_1^k|
+a_2\sum _{k=1}^N|e^k\rangle \langle g_2^k| \right) + H.c.
\label{Hami}
\end{eqnarray}
The general superposition state of an input qubit is expressed by
the form $(\alpha a_1^{\dagger }+\beta a_2^{\dagger })|0,0\rangle
=\alpha |1,0\rangle +\beta |0,1\rangle .$ The initial state
considered in \cite{SWZ,KSW} takes the following form
\begin{eqnarray}
|\Psi _{in}\rangle =\otimes _{k=1}^N|e^k\rangle
\frac {(a_1^{\dagger })^m}{\sqrt {m!}}|0,0\rangle .
\label{oinput}
\end{eqnarray}

Suppose we want to clone $M$ identical pure states $|\Phi \rangle
^{\otimes M}=(\alpha a_1^{\dagger } +\beta a_2^{\dagger
})^{\otimes N}|0,0\rangle $. It is then argued that we only need
to consider the cloning of initial state (\ref{oinput}) with the
Hamiltonian (\ref{Hami}) \cite{SWZ,KSW}. Here we present another
method. If we know how to clone the state
\begin{eqnarray}
|\Psi _{in},j\rangle=\otimes _{k=1}^M|e^k\rangle
\frac {(a_1^{\dagger })^{M-j}
(a_2^{\dagger })^{j}}{\sqrt {(M-j)!j!}}|0,0\rangle ,
\nonumber \\
j=0, 1,\cdots, M,
\end{eqnarray}
it will be straightforward to clone the $M$ identical pure states
$|\Phi \rangle ^{\otimes M}=(\alpha a_1^{\dagger } +\beta
a_2^{\dagger })^{\otimes M}|0,0\rangle $. And more interesting,
we can extend the input of the UQCM to arbitrary states in the
Bose subspace because $\frac {(a_1^{\dagger })^{i} (a_2^{\dagger
})^{j}}{\sqrt {i!j!}}|0,0\rangle , i+j=M$ constitutes a complete
orthonormal basis of the Bose subspace with $M$ qubits. We remark
here that arbitrary input states in Bose subspace also include
mixed states.

For convenience, we use the same notation as in \cite{SWZ,KSW}.
Using the Schwinger representation we denote the total angular
momentum operator as $b_rc^{\dagger }\equiv \sum
_{k=1}^N|e^k\rangle \langle g_r^k|, ~~~r=1,2$, where $c^{\dagger}
$ is the creation operator of "e-type" excitations and $b_r$ is
the annihilation operator of the ground states $g_r$, $r=1,2$.
The Hamiltonian (\ref{Hami}) then becomes as follows in terms of
harmonic-oscillator operators
\begin{eqnarray}
{\cal {H}}=\gamma (a_1b_1+a_2b_2)c^{\dagger }+H.c.
\label{Hami-osc}
\end{eqnarray}
Now we study the case of initial states containing both kinds of
oscillators $a_1^{\dagger }$ and $a_2^{\dagger }$ of $i+j$ qubits,
\begin{eqnarray}
|\Psi _{in},(i,j)\rangle &=&
\frac {(a_1^{\dagger })^{i}(a_2^{\dagger })^j(c^{\dagger })^N}
{\sqrt {i!j!N!}}|0\rangle
\nonumber \\
&=&|i_{a_1},j_{a_2}\rangle |0_{b_1},0_{b_2}\rangle |N_c\rangle
\nonumber \\
&\equiv &|i,j\rangle _a|0,0\rangle _b|N\rangle _c.
\label{g-input}
\end{eqnarray}
With the initial state (\ref{g-input}),
the time evolution of the state acts as follows\cite{KSW}
\begin{eqnarray}
&&|\Psi (t),(i,j)\rangle =e^{-iHt}|\Psi _{in},(i,j)\rangle
\nonumber \\
&=&\sum _p(-iHt)^p/p!|\Psi _{in},(i,j)\rangle
=\sum _{l=0}^Nf_l(t)|F_l,(i,j)\rangle ,
\label{time}
\end{eqnarray}
where $|\Psi _{in},(i,j)\rangle =|F_0,(i,j)\rangle$, and the state
$|F_l,(i,j)\rangle $ expresses that $i+j+l$ copies of the initial
state (\ref{g-input}) are obtained. $l$ is the number of
additional photons that have been emitted. So the output state of
this cloning machine may contains between 0 and $N$ additional
copies of the initial state (\ref{g-input}). It will always be a
superposition of $|F_l,(i,j)\rangle $ components. The probability
of finding $l$ additional copies is determined by the amplitude
$|f_l(t)|^2$ of the corresponding term. After some calculations
we find that the output with $l$ additional copies is
\begin{eqnarray}
|F_l,(i,j)\rangle =
\sum _{k=0}^l\sqrt {\frac {l!(i+j+1)!}{(i+j+l+1)!}}
\sqrt{ \frac {(i+l-k)!(j+k)!}{i!j!k!(l-k)!}}
\nonumber \\
|i+l-k,j+k\rangle _a|l-k,k\rangle _b|N-l\rangle _c.
\label{g-output}
\end{eqnarray}

The cloning transformation takes $i+j$ qubits in the form
(\ref{g-input}) as an input, and produces $i+j+l$ output qubits in
the form (\ref{g-output}). The action of the Hamiltonian
(\ref{Hami-osc}) on the state $|F_l,(i,j)\rangle $ is as follows.
\begin{eqnarray}
&&{\cal {H}}|F_l,(i,j)\rangle
\nonumber \\
&=&\gamma (\sqrt{(l+1)(N-l)(i+j+l+2)}|F_{l+1},(i,j)\rangle
\nonumber \\
&&+\sqrt{l(N-l+1)(i+j+l+1)}|F_{l-1},(i,j)\rangle ,
\nonumber \\
&&~~~~l\le l<N,
\nonumber \\
&&{\cal {H}}|F_0,(i,j)\rangle =\gamma \sqrt{N(i+j+2)}|F_1,(i,j)\rangle ,
\nonumber \\
&&{\cal {H}}|F_N,(i,j)\rangle =\gamma \sqrt{N(i+j+N+1)}|F_{N-1},(i,j)\rangle .
\nonumber \\
&&
\label{struct}
\end{eqnarray}
We remark that in the case where $i=m,j=0$, we recover the
results of \cite{KSW}.

We now consider the cloning of $M$ identical pure input states to
$L$ copies. We have
\begin{eqnarray}
&&|\Phi \rangle ^{\otimes M}=(\alpha a_1^{\dagger }
+\beta a_2^{\dagger })^{\otimes M}|0,0\rangle
\nonumber \\
&&=\sum _{j=0}^M\frac {M!}{\sqrt{(M-j)!j!}}\alpha ^{M-j}\beta ^j
\frac {(a_1^{\dagger })^{M-j}(a_2^{\dagger }) ^j}{\sqrt{(M-j)!j!}}
|0,0\rangle
\end{eqnarray}
and we already know the cloning of the basis states $\frac
{(a_1^{\dagger })^{M-j}(a_2^{\dagger }) ^j}{\sqrt{(M-j)!j!}}
|0,0\rangle $ Using the cloning transformation (\ref{g-output}),
we can obtain the $L$-copy output as
\begin{eqnarray}
&&|\Phi \rangle ^{out}=
\sum _{j=0}^M\sum _{k=0}^{L-M}
\frac {M!}{\sqrt{(M-j)!j!}}\alpha ^{M-j}\beta ^j
\nonumber \\
&&\sqrt{\frac {(L-M)!(M+1)!}{(L+1)!}}
\sqrt {\frac {(L-j-k)!(j+k)!}{(M-j)!j!k!(L-M-k)!}}
\nonumber \\
&&|L-j-k,k+j\rangle _a|L-M-k,k\rangle _b|N-(M-L)\rangle _c.
\end{eqnarray}
The density operator of the photonic output can be obtained by
tracing over the $b$ and $c$ modes. By tracing out all but one
qubit ($a$ modes) we can obtain the output reduced density
operator. The fidelity calculated therefrom is
$F=M(L+2)+L-M/L(M+2)$, which is known to be optimal. Therefore we
have shown both that the cloning transformation is universal and
optimal. A further criterion will later show that the cloning
transformation (\ref{g-output}) is also optimal and universal in
the d-level case.

We know from (\ref{g-input}) that the cloning machine allows
arbitrary mixed input states in Bose subspace. We consider an
input state of M qubits of the form
\begin{eqnarray}
\rho =\sum _{jj'}^M\alpha _{jj'}
\frac {(a_1^{\dagger })^{M-j}(a_2^{\dagger }) ^j}{\sqrt{(M-j)!j!}}
|00\rangle \langle 00|
\frac {(a_1)^{M-j'}(a_2) ^{j'}}{\sqrt{(M-j')!j'!}},
\label{general}
\end{eqnarray}
where $\alpha _{jj'}$ are arbitrary parameters. Certainly we need
$\rho $ to be a density operator. Using the cloning
transformation (\ref{g-output}), we can obtain $l$ additional
copies. It can be proved that with a general input
(\ref{general}), the transformation (\ref{g-output}) is still
optimal\cite{FMWHW}. We remark that (\ref{g-input}) even allows
the input states to be composed of different qubits if they can be
expressed by Bosonic operators, although the interpretation is
not as obvious as for identical qubits.

In the following we will present the main result of our work. We
will study the optimal photon cloning of states in a d-dimensional
Hilbert space (qudits). The atoms of the inverted cloning medium
now must have one excited state $|e\rangle $ and $d$ ($d\ge 2$)
ground states $|g_n\rangle , n=1, 2,\cdots, d$, where again each
transition is coupled to a different degree of freedom of the
photons $a_n$. Similar to the qubit case, we denote by
$b_rc^{\dagger }\equiv \sum _{k=1}^N|e^k\rangle \langle g_r^k|,
~~~r=1,\cdots, d$. The Hamiltonian of
the cloning system in terms of harmonic-oscillator operators is
written as \cite{KSW}
\begin{eqnarray}
{\cal {H}}_d=\gamma (a_1b_1+\cdots +a_db_d)+H.c.
\label{d-Hami}
\end{eqnarray}
Again we consider general initial states in the Bose subspace
\begin{eqnarray}
|\Psi _{in},\vec {j}\rangle =
\prod _{i=1}^d\frac {(a_i^{\dagger })^{j_i}}{\sqrt{j_i!}}
\frac {(c^{\dagger })^N}{\sqrt{N!}}|0\rangle
\equiv |\vec{j}\rangle _a|\vec{0}\rangle _b|N\rangle _c,
\label{d-input}
\end{eqnarray}
where $\vec {j}=(j_1,j_2,\cdots,j_d)$.
There are still $N$ excited states
in the initial state, so the number of additional copies is
restricted by N. We remark that the initial states (\ref{d-input})
to be cloned span arbitrary states in the Bose subspace and
constitute an orthonormal basis.
One can see easily that the time evolution of states for
qudits is the same as the one for qubits presented in
(\ref{time}). That means the probability to obtain $l$ additional
copies is still $|f_l(t)|^2$. We denote
$|F_0,\vec{j}\rangle \equiv |\Psi _{in},\vec{j}\rangle $, $\sum
_{i}j_i=M$. The output of the cloning system with $l$
additional copies of the initial state (\ref{d-input}) can be
calculated as
\begin{eqnarray}
|F_l,\vec{j}\rangle &=&
\sum _{k_i}^l\sqrt{ \frac {(M+d-1)!l!}{(M+l+d-1)!}}
\prod _{i=1}^d\sqrt{ \frac {(k_i+j_i)!}{k_i!j_i!}}
\nonumber \\
&&|\vec{j}+\vec{k}\rangle _a|\vec{k}\rangle _b|N-l\rangle _c,
\label{d-output}
\end{eqnarray}
where the summation $\sum _{k_i}^l$ means taking the sum over all
variables under the condition $\sum _i^dk_i=l$.

It is very interesting that the cloning
transformation (\ref{d-output}) with input (\ref{d-input}) is
completely determined by the interaction Hamiltonian
(\ref{d-Hami}). Given different input states to be cloned, the
action of the Hamiltonian on the initial states will produce the
corresponding cloning output. That means the procedure of quantum
cloning is completely controlled by the Hamiltonian as showed
by the following calculations
\begin{eqnarray}
&&{\cal {H}}_d|F_l,\vec{j}\rangle
=\gamma (\sqrt{(l+1)(N-l)(M+l+d)}|F_{l+1},\vec{j}\rangle
\nonumber \\
&&+\sqrt{l(N-l+1)(M+l+d-1)}|F_{l-1},\vec{j}\rangle ,
\nonumber \\
&&~~~~l\le l<N,
\nonumber \\
&&{\cal {H}}_d|F_0,\vec{j}\rangle =\gamma \sqrt{N(M+d)}|F_1,\vec{j}\rangle ,
\nonumber \\
&&{\cal {H}}_d|F_N,\vec{j}\rangle =\gamma \sqrt{N(M+N+d-1)}|F_{N-1},\vec{j}
\rangle .
\label{struct1}
\end{eqnarray}

Now we see how to clone $M$ identical qudits to $M+l\equiv L$
copies. An arbitrary qudit takes the form $|\Psi \rangle =\sum
_{i=1}^dx_ia_i^{\dagger }|\vec{0}\rangle $, with $\sum
_{i=1}^d|x_i|^2=1$. The state of $M$ identical qudits to be cloned
can be expressed as follows
\begin{eqnarray}
|\Psi \rangle ^{\otimes M}&=&
(\sum _{i=1}^dx_ia_i^{\dagger })^{\otimes M}|\vec{0}\rangle
\nonumber \\
&=&M!\sum _{j_i}^M
\prod _{i=1}^d
\frac {x_i^{j_i}}
{\sqrt{j_i!}}
\frac {(a_i^{\dagger })^{j_i}}
{\sqrt {j_i!}}|\vec{0}\rangle .
\end{eqnarray}
Consider that we intend to clone this state in the system with
$N$ atoms in the excited state $|e\rangle $. This implies that the
number of additional copies is restricted by $N$. With the help
of the cloning transformation (\ref{d-output}), we find that the
output of $L$ copies of $M$ identical input qudits has the
following form
\begin{eqnarray}
|\Psi \rangle ^{out}
&=&M!\sum _{j_i}^M
\sum _{k_i}^l\sqrt{ \frac {(M+d-1)!l!}{(L+d-1)!}}
\nonumber \\
&&\times \prod _{i=1}^d
\frac {x_i^{j_i}}
{j_i!}
\sqrt{ \frac {(k_i+j_i)!}{k_i!}}
|\vec{j}+\vec{k}\rangle _a|\vec{k}\rangle _b,
\label{out}
\end{eqnarray}
where we omit the type $c$ modes which count the number of clones
that the system has produced. We can calculate the fidelity of
the cloning transformation to be
\begin{eqnarray}
F=\langle \Psi |\rho ^{out}_{red.}|\Psi \rangle
=\frac {M(L+d)+{L-M}}{L(M+d)},
\end{eqnarray}
where $\rho ^{out}_{red.}$ means taking the trace over the ancilla
modes, i.e. $b$-type modes, and over all but one $a$ mode of
$\rho ^{out}=|\Psi \rangle ^{out~out}\langle \Psi |$. This
fidelity is the optimal fidelity for identical pure input states
in $d$-dimensional Hilbert space\cite{W,KW}. Also, obviously the
cloning transformation is universal.

Next, instead of the fidelity between input and output reduced
density operators of a single qudit, we use the fidelity between
the output of $L$ qudits and $L$ identical pure qudits as the
measure of quality for the cloning transformation
(\ref{d-output}). The definition of this fidelity takes the
following form:
\begin{eqnarray}
{\cal {F}}=^{L\otimes }\langle \Psi |
Tr_b\{ |\Psi \rangle ^{out~out}\langle \Psi |\}
|\Psi \rangle ^{\otimes L}
\end{eqnarray}
With the help of the result (\ref{out}), and considering the
normalization, we find that
\begin{eqnarray}
{\cal {F}}=\frac {L!(M+d-1)!}{M!(L+d-1)!}.
\end{eqnarray}
This is the optimal cloning fidelity for identical pure input
states\cite{W}. We remark that optimal cloning of pure states was
studied by Werner et al.\cite{W,KW} using completely positive (CP)
maps realized by symmetric projection operators. In this paper,
quantum cloning (\ref{d-output}) is obtained from the Hamiltonian
and is consequently realized by a unitary transformation. Thus we
show that for both density operator and reduced density operator,
the fidelities of the cloning transformation in (\ref{d-output})
are optimal for identical pure input states. Next, we study the
cloning of arbitrary symmetric mixed states of d-level quantum
systems.

As in the qubit case, relation (\ref{d-input}) also admits
arbitrary mixed qudits in Bose subspace as inputs. The most
general input states can take the form
\begin{eqnarray}
\rho =\sum _{\vec{j}\vec{j'}}^M
\alpha _{\vec{j}\vec{j'}}
|\vec{j}\rangle _{a~a}
\langle \vec{j'}|
\end{eqnarray}
with arbitrary parameters $\alpha _{\vec{j}\vec{j'}}$ under the
restriction that $\rho $ is a density operator. We can
prove\cite{FMWHW} that the cloning transformation
(\ref{d-output}) is the optimal cloning transformation, i.e., the
shrinking factor between the input and output reduced density
operators of a single qudit achieves its upper bound.

Formally the cloning of d-level quantum systems can be optimally
realized in the atomic system presented above. A pratical
difficulty arises because the proposal assumes $d$ distinct
photonic modes. Obviously the photon's polarization provides only
two of these modes. The fact that it would be very difficult, if
possible at all, to achieve the degeneracy of the corresponding
atomic transitions leads to the idea that one could use various
longitudinal (frequency) modes of the photon field as a basis.
Still, we are aware of the difficulty of identifying a level
scheme where the couplings to the light field are all equally
strong. But even if the experimental realization of our scheme
seems difficult, it might still be easier than to set up the
equivalent quantum network and it shows that generalized quantum
cloning is very naturally embedded in the interaction of atoms
and photons.

Very recently an experiment by Nuclear Magnetic Resonance to
implement the one to two UQCM as a quantum network has been
reported\cite{CJ}. Another experiment which demonstrates the onset
of laser-like action for entangled photons\cite{LHB} has been
extended to implement cloning of photons in parametric
down-conversion\cite{LHSB}. The experiment showed very good
agreement with the theoretical predictions.

{\it Acknowlegements:} G. W. would like to acknowledge the
University of Vienna for granting him a leave during which this
work was prepared. We thank X.B.Wang for useful discussions.


\begin{thebibliography}{99}
\bibitem{WZ}W. K. Wootters, and W. H. Zurek, Nature (London){\bf 299},
802(1982).
\bibitem{BCFJ}H. Barnum, C. Caves, C. A. Fuchs, and B. Schumacher,
Phys. Rev. Lett. {\bf 76}, 2818(1996).
\bibitem{KI} M. Koashi, and N. Imoto, Phys. Rev. Lett. {\bf 81}, 4264 (1998).
\bibitem{BH}V. Bu\v{z}ek, and M. Hillery, Phys. Rev. A{\bf 54}, 1844(1996).
\bibitem{BDEF}D. Bru\ss , D. DiVincenzo, A. Ekert, C. A. Fuchs,
C. Macchiavello, and J. A. Smolin, Phys. Rev. A{\bf 57},
2368(1998).
\bibitem{GM}N. Gisin, and S. Massar, Phys. Rev. Lett. {\bf 79},2153(1997).
\bibitem{W}R. F. Werner, Phys. Rev. A{\bf 58}, 1827(1998).
\bibitem{KW}M. Keyl and R. F. Werner, J. Math. Phys. {\bf 40}, 3283 (1999).
\bibitem{BH1} V. Bu\v{z}ek, and H. Hillery, Phys. Rev. Lett. {\bf 81},
5003 (1998).
\bibitem{FMW}H. Fan, K. Matsumoto, and M. Wadati, Phys. Rev. A{\bf 64},
064301 (2001).
\bibitem{FMWHW}H. Fan, K. Matsumoto, X. B. Wang, H. Imai, and M. Wadati,
quant-ph/0107113.
\bibitem{DBE}R. Derka, V.Bu\v{z}ek, and A.Ekert, Phys. Rev. lett. {\bf 80},
1571 (1998).
\bibitem{BEM}D. Bru\ss ,A. Ekert, and C. Macchiavello,
Phys. Rev. Lett. {\bf 81}, 2598(1998).
\bibitem{BBHB}V. Bu\v{z}ek, S. L. Braunstein, M. Hillery, and
D. Bru\ss , Phys. Rev. A{\bf 56}, 3446 (1997).
\bibitem{SWZ} C. Simon, G. Weihs, and A. Zeilinger, Phys. Rev. Lett. {\bf 84},
2993 (2000).
\bibitem{KSW}J. Kempe, C. Simon, and G. Weihs, Phys. Rev. A{\bf 62},
032302 (2000).
\bibitem{GP}N.Gisin and S.Popescu, Phys. Rev. Lett. {\bf 83}, 432 (1999).
\bibitem{FIMC}J.Fiurasek, S.Iblisdir, S.Massar, and N.J.Cerf,
quant-ph/0110016.
\bibitem{CJ}H.K.Cummins, et al, quant-ph/0111098.
\bibitem{LHB}A.Lamas-Linares, J.C.Howell, and D.Bouwmeester,
Nature {\bf 412}, 887 (2001).
\bibitem{LHSB}A. Lamas-Linares, J. C. Howell, C. Simon, and D.
Bouwmeester, submitted for publication (2001).
\end{thebibliography}
\end{document}